# Degenerate wave-like solutions to the Dirac equation for massive particles

Georgios N. Tsigaridas[1,*], Aristides I. Kechriniotis[2], Christos A. Tsonos[2] and Konstantinos K. Delibasis[3]

[1]Department of Physics, School of Applied Mathematical and Physical Sciences, National Technical University of Athens, GR-15780 Zografou Athens, Greece

[2]Department of Physics, University of Thessaly, GR-35100 Lamia, Greece

[3]Department of Computer Science and Biomedical Informatics, University of Thessaly, GR-35131 Lamia, Greece

[*]Corresponding Author. E-mail: gtsig@mail.ntua.gr

**Abstract**

In this work we provide a novel class of degenerate solutions to the Dirac equation for massive particles, where the rotation of the spin of the particles is synchronized with the rotation of the magnetic field of the wave-like electromagnetic fields corresponding to these solutions. We show that the state of the particles does not depend on the intensity of the electromagnetic fields but only on their frequency, which is proportional to the mass of the particles and lies in the region of Gamma/X-rays for typical elementary charged particles, such as electrons and protons. These novel theoretical results could play an important role in plasma physics, astrophysics, and other fields of physics, involving the interaction of charged particles with high energy photons.

1. **Introduction**

In a recent article [1] we have shown that all solutions to the Dirac equation

$$i\gamma^\mu \partial_\mu \Psi + a_\mu \gamma^\mu \Psi - m\Psi = 0 \qquad (1)$$

satisfying the conditions $\Psi^\dagger \gamma \Psi = 0$ and $\Psi^T \gamma^2 \Psi \neq 0$, where $\gamma^\mu$ are the standard Dirac matrices and $\gamma = \gamma^0 + i\gamma^1\gamma^2\gamma^3$ are degenerate, corresponding to an infinite number of electromagnetic 4-potentials $A_\mu$, explicitly calculated in Theorem 5.4 in [1]. In the Dirac equation $m$ and $q$ are the mass and the electric charge of the particle, respectively, and $a_\mu = qA_\mu$. It should also be noted that Eq. (1) is written in natural units, where both the speed of light in vacuum $c$ and the reduced Planck constant $\hbar$ are equal to one. Furthermore, in [1] we have shown that all solutions to the Weyl equation are degenerate, corresponding to an infinite number of electromagnetic 4-potentials, explicitly calculated in Theorem 3.1 in [1]. Some very interesting properties of Weyl particles, mainly regarding their control and localization, are discussed in [2, 3].

As it has also been shown in [1], the degenerate solutions in the case of free Dirac particles correspond to massless particles, except for particle-antiparticle pairs. However, the net charge of the particle – antiparticle pair is zero, and consequently the degeneracy is not particularly meaningful from a practical point of view. Additionally, in a recent work [4] we have shown that degenerate solutions for massive particles can exist in potential barriers. However, these solutions involve real exponential terms and consequently, they cannot describe the state of particles in free space. Finally, it should be mentioned that in [5] we provide a general method for obtaining degenerate solutions to the Dirac and Weyl equations, both for massive and massless particles.

## 2. Degenerate wave-like solutions to the Dirac equation and the corresponding 4-potentials

In this work we investigate the existence of degenerate solutions for massive Dirac particles involving only complex exponential terms, which are well defined throughout space and time. For this purpose, we have used the following general form of degenerate spinors

$$\Psi = \begin{pmatrix} \cos\alpha \exp(ih_1) \\ \sin\alpha \exp(ih_2) \\ \cos\beta \exp(ih_1) \\ \sin\beta \exp(ih_2) \end{pmatrix} \quad (2)$$

where $h_1, h_2$ are arbitrary real functions of the spatial coordinates and time and $\alpha, \beta$ are real constants. Using the above ansatz and requiring $\Psi$ to be solution to the Dirac equation for real 4-potentials it is found that all spinors of the form

$$\Psi = c_1 \exp(ih) \begin{pmatrix} \cos\alpha \\ \sin\alpha \exp(id) \\ \cos\beta \\ \sin\beta \exp(id) \end{pmatrix} \quad (3)$$

where $c_1$ is an arbitrary complex constant, $h$ an arbitrary real function of the spatial coordinates and time and

$$d = \frac{4m[t - z\cos(\alpha+\beta)]}{\cos(2\alpha) - \cos(2\beta)} \quad (4)$$

are degenerate, satisfying the Dirac equation for the following 4-potentials:

$$\begin{pmatrix} a_0 \\ a_1 \\ a_2 \\ a_3 \end{pmatrix} = \begin{pmatrix} -\dfrac{\tan(\alpha+\beta)\left[2\sin(\alpha-\beta)\left(\dfrac{\partial h}{\partial z}-g\right)+(\sin(2\alpha)-\sin(2\beta))\dfrac{\partial h}{\partial t}+m(\sin(2\alpha)+\sin(2\beta))\right]}{\cos(2\alpha)-\cos(2\beta)} \\ -\sec(\alpha+\beta)\cos d\left[\sin(\alpha+\beta)\left(\dfrac{\partial h}{\partial z}-g\right)+2m\cos\alpha\cos\beta\csc(\alpha-\beta)\right]+\dfrac{\partial h}{\partial x} \\ \dfrac{1}{2}\csc(\alpha-\beta)\sec(\alpha+\beta)\sin d\left[(\cos(2\alpha)-\cos(2\beta))\left(\dfrac{\partial h}{\partial z}-g\right)-4m\cos\alpha\cos\beta\right]+\dfrac{\partial h}{\partial y} \\ g \end{pmatrix}$$

(5)

where $g$ is an arbitrary real function of the spatial coordinates and time. In the above expressions we have supposed that $\cos(2\alpha)-\cos(2\beta)\neq 0$, $\cos(\alpha+\beta)\neq 0$, $\sin(\alpha-\beta)\neq 0$, implying that

$$\alpha\pm\beta \neq n\pi \text{ and } \alpha+\beta \neq n\pi+\pi/2,\ n\in\mathbb{Z} \qquad (6)$$

An important characteristic of the degenerate spinors given by Eq. (3) is that they can describe particles of any mass, including massless particles. In addition, the above solutions correspond to particles in non-localized states, which can exist throughout space and time without any restriction. In contrast, the degenerate solutions in [4] describe particles existing only in classically forbidden regions.

Furthermore, setting $g = (\partial h/\partial z)$, the 4-potentials given by Eq. (5) take the simplified form

$$\begin{pmatrix} a_0 \\ a_1 \\ a_2 \\ a_3 \end{pmatrix} = \begin{pmatrix} -\dfrac{\tan(\alpha+\beta)\left[(\sin(2\alpha)-\sin(2\beta))\dfrac{\partial h}{\partial t}+m(\sin(2\alpha)+\sin(2\beta))\right]}{\cos(2\alpha)-\cos(2\beta)} \\ -2m\cos\alpha\cos\beta\csc(\alpha-\beta)\sec(\alpha+\beta)\cos d+\dfrac{\partial h}{\partial x} \\ -2m\cos\alpha\cos\beta\csc(\alpha-\beta)\sec(\alpha+\beta)\sin d+\dfrac{\partial h}{\partial y} \\ \dfrac{\partial h}{\partial z} \end{pmatrix} \qquad (7)$$

An interesting feature of the 4-potentials given by Eq. (7) is that they become zero in the case of

$$\frac{\partial h}{\partial x} = \frac{\partial h}{\partial y} = \frac{\partial h}{\partial z} = 0 \qquad (8)$$

$$\alpha = n\pi + \frac{\pi}{2} \text{ or } \beta = n\pi + \frac{\pi}{2},\ n\in\mathbb{Z} \qquad (9)$$

and

$$\frac{\partial h}{\partial t} = -m \frac{\sin(2\alpha) + \sin(2\beta)}{\sin(2\alpha) - \sin(2\beta)} \tag{10}$$

In the above it is assumed that $\sin(2\alpha) - \sin(2\beta) \neq 0$, which is true if the conditions described by Eq. (6) are valid. Thus, the degenerate spinors given by Eq. (3), in the case of the conditions described by equations (8) - (10), are solutions to the Dirac equation for zero 4-potential and consequently zero electromagnetic field. As an example, we consider the spinors

$$\Psi_0 = c_1 \exp(imt) \begin{pmatrix} 0 \\ \exp(id) \\ \cos\beta \\ \sin\beta \exp(id) \end{pmatrix}, \quad \Psi'_0 = c_1 \exp(-imt) \begin{pmatrix} \cos\alpha \\ \sin\alpha \exp(id) \\ 0 \\ \exp(id) \end{pmatrix} \tag{11}$$

corresponding to zero electromagnetic 4-potential and field.

### 3. The electromagnetic fields corresponding to the degenerate solutions and some important remarks

The electromagnetic fields (in Gaussian units) corresponding to the 4-potentials given by Eq. (7) are given by the following formulae [6, 7]:

$$\mathbf{E} = -\nabla U - \frac{\partial \mathbf{A}}{\partial t} \tag{12}$$
$$= 4m^2 \cos\alpha \cos\beta \csc^2(\alpha - \beta) \csc(\alpha + \beta) \sec(\alpha + \beta)(-\sin d\, \mathbf{i} + \cos d\, \mathbf{j})$$

$$\mathbf{B} = \nabla \times \mathbf{A} = -4m^2 \cos\alpha \cos\beta \csc^2(\alpha - \beta) \csc(\alpha + \beta)(\cos d\, \mathbf{i} + \sin d\, \mathbf{j}) \tag{13}$$

where $U = a_0/q$ is the electric potential and $\mathbf{A} = -(1/q)(a_1\mathbf{i} + a_2\mathbf{j} + a_3\mathbf{k})$ is the magnetic vector potential. The above equations are expressed in the natural system of units, where $\hbar = c = 1$.

An interesting remark is that the electromagnetic fields given by equations (12), (13) resemble a circularly polarized plane wave propagating along the +z-direction with Poynting vector

$$\mathbf{S} = \frac{1}{4\pi} \mathbf{E} \times \mathbf{B} = \frac{4m^4}{\pi} \cos^2\alpha \cos^2\beta \csc^4(\alpha - \beta) \csc^2(\alpha + \beta) \sec(\alpha + \beta) \mathbf{k} \tag{14}$$

In addition, according to Theorem 5.4 in [1], the spinors given by Eq. (3) will also be solutions to the Dirac equation for an infinite number of 4-potentials, given by the formula

$$b_\mu = a_\mu + s\kappa_\mu \tag{15}$$

where

$$(\kappa_0, \kappa_1, \kappa_2, \kappa_3) = \left(1, -\frac{\Psi^T \gamma^0 \gamma^1 \gamma^2 \Psi}{\Psi^T \gamma^2 \Psi}, -\frac{\Psi^T \gamma^0 \Psi}{\Psi^T \gamma^2 \Psi}, \frac{\Psi^T \gamma^0 \gamma^2 \gamma^3 \Psi}{\Psi^T \gamma^2 \Psi}\right) \tag{16}$$
$$= (1, -\sin(\alpha+\beta)\cos d, -\sin(\alpha+\beta)\sin d, -\cos(\alpha+\beta))$$

and $s$ is an arbitrary real function of the spatial coordinates and time.

The electromagnetic fields corresponding to the 4-potentials $b_\mu - a_\mu = \kappa_\mu s$ are the following:

$$\begin{aligned}
\mathbf{E}_s = &-\left(2ms_q \csc(\alpha-\beta)\sin d + \sin(\alpha+\beta)\cos d \frac{\partial s_q}{\partial t} + \frac{\partial s_q}{\partial x}\right)\mathbf{i} \\
&+ \left(2ms_q \csc(\alpha-\beta)\cos d - \sin(\alpha+\beta)\sin d \frac{\partial s_q}{\partial t} - \frac{\partial s_q}{\partial y}\right)\mathbf{j} \\
&- \left(\cos(\alpha+\beta)\frac{\partial s_q}{\partial t} + \frac{\partial s_q}{\partial z}\right)\mathbf{k}
\end{aligned} \tag{17}$$

$$\begin{aligned}
\mathbf{B}_s = &-\left(\sin(\alpha+\beta)\sin d \frac{\partial s_q}{\partial z} + \cos(\alpha+\beta)\left(2ms_q \csc(\alpha-\beta)\cos d - \frac{\partial s_q}{\partial y}\right)\right)\mathbf{i} \\
&+ \left(\sin(\alpha+\beta)\cos d \frac{\partial s_q}{\partial z} - \cos(\alpha+\beta)\left(2ms_q \csc(\alpha-\beta)\sin d + \frac{\partial s_q}{\partial x}\right)\right)\mathbf{j} \\
&+ \sin(\alpha+\beta)\left(-\cos d \frac{\partial s_q}{\partial y} + \sin d \frac{\partial s_q}{\partial x}\right)\mathbf{k}
\end{aligned} \tag{18}$$

In the above, $s_q = s/q$ is an arbitrary real function of the spatial coordinates and time. Consequently, particles described by the degenerate spinors given by Eq. (3) have the remarkable property to exist in the same quantum state in the wide variety of electromagnetic fields described by equations (12), (13) and (17), (18). It should also be noted that the factor $d$, given by Eq. (4), is also involved in the electromagnetic fields $\mathbf{E}_s, \mathbf{B}_s$ and consequently, they are expected to have similar properties with the fields described by equations (12), (13).

As an example, we consider the case of constant function $s$. Then, the electromagnetic fields corresponding to the 4-potentias $b_\mu$ take the following form:

$$\begin{aligned}
\mathbf{E}_{t,w} = &\,2m\csc(\alpha-\beta)\left(2m\cos\alpha\cos\beta\csc(\alpha-\beta)\csc(\alpha+\beta)\sec(\alpha+\beta) + s_q\right) \\
&\times(-\sin d\,\mathbf{i} + \cos d\,\mathbf{j})
\end{aligned} \tag{19}$$

$$\mathbf{B}_{t,w} = -2m\csc(\alpha-\beta)\big(2m\cos\alpha\cos\beta\csc^2(\alpha-\beta)\csc(\alpha+\beta)+s_q\cos(\alpha+\beta)\big) \quad (20)$$
$$\times(\cos d\ \mathbf{i}+\sin d\ \mathbf{j})$$

having the same spatiotemporal dependence with the electromagnetic fields given by equations (12), (13). This practically means that the state of the particles does not depend on the magnitude of the wave-like electromagnetic fields, but only on their spatial and temporal dependence, given by the function

$$d = \frac{4m[t-z\cos(\alpha+\beta)]}{\cos(2\alpha)-\cos(2\beta)} = \omega_d t - k_d z \quad (21)$$

where

$$\omega_d = \frac{4m}{\cos(2\alpha)-\cos(2\beta)} \quad (22)$$

and

$$k_d = \frac{4m\cos(\alpha+\beta)}{\cos(2\alpha)-\cos(2\beta)} \quad (23)$$

are constants related to the angular frequency and the wavenumber, respectively. It should be noted that the phase velocity

$$\upsilon_{ph} = \frac{\omega_d}{k_d} = \sec(\alpha+\beta) \quad (24)$$

is higher than the speed of light ($c=1$ in natural units). However, this does not violate the special theory of relativity since a sinusoidal wave with a unique frequency does not transmit any information. It is reminded that the phase velocity of an electromagnetic wave propagating through a medium can exceed the speed of light in vacuum, as it happens in most glasses at X-ray frequencies [8] and in unmagnetized plasmas [9].

Another interesting remark is that the frequency of these wave-like fields depends on the mass of the particles. In more detail, in S.I. units, the factor $4m$ in Eq. (20) becomes $4mc^2/\hbar$ and consequently the frequency of the oscillation is given by the formula

$$f_d(\text{S.I.}) = \frac{\omega_d(\text{S.I.})}{2\pi} = \frac{4mc^2}{h}\frac{1}{\cos(2\alpha)-\cos(2\beta)} \quad (25)$$

For example, in the case of electrons $(m_e = 9.109\times10^{-31}\,Kg)$ the frequency of the oscillation becomes

$$f_d(\text{S.I.}) = \frac{4.95 \times 10^{20}}{\cos(2\alpha) - \cos(2\beta)} Hz \tag{26}$$

corresponding to photons with energy higher than 2.05 MeV, in the region of Gamma/X-rays. Furthermore, in the case of heavier particles, e.g., protons $(m_p = 1.673 \times 10^{-27} kg)$, the oscillation frequency takes much higher values, above $9.09 \times 10^{23} Hz$, corresponding to photons with extremely high energy, higher than 3.75 GeV.

Consequently, the innovative results presented in this article are expected to be applied in many fields of physics involving the interaction of high energy photons with charged particles, especially in the framework of plasma physics and astrophysics. For example, they could be used to explain several phenomena regarding the interaction of cosmic rays with ionized gas clouds.

It should also be noted that, setting $\cos(2\alpha) - \cos(2\beta) = 2/n$, $n \in \mathbb{Z}$, in Eq. (25), the energy of the photons becomes exactly equal to the one required for the production of $n$ particle – antiparticle pairs. Consequently, the degenerate spinors given by Eq. (3) could also be related to the Schwinger effect [10-16].

In addition, it is particularly interesting to study the spin of the particles described by the degenerate spinors given by Eq. (3). Specifically, the expected values of the projections of the spin of the particles along the x, y, and z-axes are given by the following formulae [17, 18]:

$$S_x = \frac{i}{2} \Psi^\dagger \gamma^2 \gamma^3 \Psi = \frac{|c_1|^2}{2} (\sin(2\alpha) + \sin(2\beta)) \cos d \tag{27}$$

$$S_y = \frac{i}{2} \Psi^\dagger \gamma^3 \gamma^1 \Psi = \frac{|c_1|^2}{2} (\sin(2\alpha) + \sin(2\beta)) \sin d \tag{28}$$

$$S_z = \frac{i}{2} \Psi^\dagger \gamma^1 \gamma^2 \Psi = \frac{|c_1|^2}{2} (\cos(2\alpha) + \cos(2\beta)) \tag{29}$$

From the above expressions it becomes clear that the expected value of the projection of the spin of the particles on the x-y plane rotates in synchronization with the magnetic field of the wave-like electromagnetic fields given by equations (12), (13) and (17), (18). Therefore, the synchronization between the spin of the particles and the wave-like electromagnetic fields can be regarded as a key-feature of the degenerate solutions given by Eq. (3). Furthermore, under the conditions described by equations (8) – (10), the rotation of the spin of the particles occurs even in the absence of an electromagnetic field. Thus, it can be considered that, in the case of degenerate solutions, the electromagnetic fields should be synchronized to the rotation of the spin of the particles and not the opposite.

## 4. Conclusions

In conclusion, we have provided a novel class of degenerate solutions to the Dirac equation for massive particles, where the key feature is the synchronization between the rotation of the spin of the particles and the magnetic field of the wave-like electromagnetic fields corresponding to these solutions. We have shown that the frequency of these wave-like electromagnetic fields depends on the mass of the particles and lies in the region of Gamma/X-rays for typical subatomic particles, such as electrons, protons, etc. Another interesting characteristic of these fields is that their phase velocity is greater than the speed of light in vacuum, which does not violate the special theory of relativity, since a sinusoidal wave with a single frequency does not transmit any information. Finally, it should be mentioned that the results presented in this article are expected to play an important role in several fields of physics, where the interaction of high energy photons with charged particles is involved, such as in the field of astrophysics, where they could provide novel insight regarding the interaction of cosmic rays with ionized gas clouds.